\documentclass[aps,prx,twocolumn,floatfix,longbibliography,nofootinbib,superscriptaddress]{revtex4-1}
\usepackage[utf8]{inputenc}
\usepackage{natbib}
\usepackage{graphicx}
\usepackage{xcolor}
\usepackage[tbtags]{amsmath}
\usepackage[colorlinks, linkcolor=blue]{hyperref}
\hypersetup{colorlinks,allcolors=black}
\usepackage{amssymb}
\usepackage{gensymb}
\usepackage{float}
\usepackage{amsmath}
\usepackage{tabularx,graphicx}
\usepackage{epstopdf}
\usepackage{latexsym}
\usepackage{color, colortbl}
\usepackage{psfrag}
\usepackage{bbm}
\usepackage{bm}
\usepackage{titlesec}
\usepackage{dsfont}
\usepackage{feynmp}
\usepackage{slashed}
\usepackage{multirow}
\usepackage{ulem}
\newcommand{\be}{\begin{equation}}
\newcommand{\ee}{\end{equation}}
\newcommand{\beq}{\begin{eqnarray}}
\newcommand{\eeq}{\end{eqnarray}}
\newcommand{\ba}{\[\begin{aligned}}
\newcommand{\ea}{\end{aligned}\]}

\newcommand{\la}{\langle}
\newcommand{\ra}{\rangle}

\renewcommand{\vec}[1]{{\bf #1}}
\renewcommand{\hat}[1]{{\bf {\widehat #1}}}
\renewcommand{\phi}{\varphi}
\renewcommand{\epsilon}{\varepsilon}

\def\nn{\nonumber}

\renewcommand{\vec}[1]{\boldsymbol{#1}}

\def \br{{\bf r}}

\def \bk{{\bf k}}

\def \ve{{\varepsilon}}
\def \l{{\ell}}

\def \D{{\Delta}}

\def \k{{\kappa}}

\def \nd{{^{\vphantom{\dagger}}}}

\def \r {\vec{r}}
\def \k {\vec{k}}

\def \ra{{\rangle}}
\def \la{{\langle}}
\def \tn{\textnormal}

\def \ba{\begin{align*}}
\def \ea{\end{align*}}

\newcounter{indice}

\begin{document}
\title{
Heuristic bounds on superconductivity and how to exceed them}
\author{Johannes S. Hofmann}
\affiliation{Department of Condensed Matter Physics, Weizmann Institute of Science, Rehovot, 76100, Israel.}
\author{Debanjan Chowdhury}
\email{debanjanchowdhury@cornell.edu}
\affiliation{Department of Physics, Cornell University, Ithaca, New York 14853, USA.}
\author{Steven A. Kivelson}
\email{kivelson@stanford.edu}
\affiliation{Department of Physics, Stanford University, Stanford, California 94305, USA.}
\author{Erez Berg}
\email{erez.berg@weizmann.ac.il}
\affiliation{Department of Condensed Matter Physics, Weizmann Institute of Science, Rehovot, 76100, Israel.}
\begin{abstract}
What limits the value of the superconducting transition temperature ($T_c$) is a question of great fundamental and practical importance. 
Various heuristic upper bounds on $T_c$ have been proposed, expressed as fractions of the Fermi temperature, $T_F$, the zero-temperature superfluid stiffness, $\rho_s(0)$, 
or a characteristic Debye frequency, $\omega_0$. We
show that while these bounds are physically motivated and are
certainly useful in many relevant situations, none of them serve as a fundamental bound on $T_c$. To demonstrate this, we provide explicit models where $T_c/T_F$ (with an appropriately defined $T_F$), $T_c/\rho_s(0)$, and $T_c/\omega_0$ are unbounded. 
\end{abstract}
\maketitle
\section{Introduction}
\label{sec:Intro}

While superconducting transition temperatures are non-universal properties, and hence not generally amenable to a simple theoretical analysis, understanding 
what physics determines $T_c$  
is of self-evident importance. 
One approach to this challenge is to focus on a key physical process that contributes to the development of superconducting order, and to formulate an upper bound -- either rigorous or heuristic -- on $T_c$~\cite{McMillan1968,McMillan1968,cohen1972comments,ginzburg1991,emery1995importance,Hazra18,Esterlis2018,esterlis2018bound}.  

In this paper, we examine three proposed bounds on $T_c$ that are expressed as a fraction of a measurable physical quantity of a given system: an appropriately defined Fermi temperature, a characteristic phonon frequency, or the zero-temperature superfluid phase stiffness. While these putative bounds are physically motivated, and provide valuable intuition in many cases of practical importance, we show by explicit counter-examples that they can be violated by an arbitrary amount. 
In addition to the fundamental importance of these results, we hope they suggest routes to further optimize $T_c$.

\section{Summary of results}
\label{sec:summary}
We briefly summarize our key results here:  
\begin{enumerate}
    \item 
The notion of an upper bound on $T_c$ in terms of an appropriately defined Fermi energy comes from the fact that, in many situations, as $E_F\rightarrow 0$, the electrons have no kinetic energy. Thus, in this limit, the superfluid stiffness must seemingly go to zero. What sets  $T_c$ in the limit of small $E_F$ is pertinent to moir\'e superconductors~\cite{Cao2018,AY19,Efetov19,SNP20,PJH21,PK21}, where the bands can be tuned to be narrow. 
To make this question 
precise, we must define  
$E_F$ in a strongly interacting system. We propose two such definitions of $E_F$, in terms of (i) the difference in the chemical potential between a system with a given density of electrons and a system with a vanishing density, or (ii) the energy dispersion of a single electron added to the empty system. 

We show that there is no general bound on $T_c/E_F$ by either definition, by studying two explicit models. In the first model, a flat band is separated by an energy gap from a broad band with pair-hopping interaction between the two. The second model consists of a pair of 
perfectly flat bands with an on-site electron-electron attraction. 
We show explicitly that the first model violates any putative $T_c/E_F$ bound when using the first definition of $E_F$ above, while the second model violates the bound using either definition of $E_F$. Both models 
have been defined on a two-dimensional lattice for convenience\footnote{ In the context of two-dimensional systems, we identify $T_c$ as the Berezinskii-Kosterlitz-Thouless (BKT) transition temperature.},  
but generalized versions of the same models in any $D>1$ can be easily seen to exhibit qualitatively similar behaviors.

\item In two-dimensional systems where $T_c$ is limited by phase fluctuations, an intuitive bound on $T_c$ is given in terms of the zero-temperature phase stiffness, $\rho_s(0)$. This comes from the relation~\cite{NK77} $T_c = \pi\rho_s(T^-_c)/2$, and the (often physically reasonable) assumption that $\rho_s(T)$ is a decreasing function of $T$, and hence $\rho_s(0)\ge\rho_s(T)$. 

We construct an explicit counter-example in a two-band model of bosons (or, equivalently, tightly-bound Cooper pairs), where $\rho_s(0)$ can be made arbitrarily smaller than $T_c$. In this model, $\rho_s(0)$ can even vanish while $T_c>0$, implying that there is a reentrant transition into the non-superconducting state below $T_c$. 

\item In electron-phonon superconductors, a heuristic bound on $T_c/\omega_0$ ($\omega_0$ being the characteristic phonon frequency) was proposed~\cite{Esterlis2018,esterlis2018bound,chubukov2020eliashberg}. The reasoning behind this bound is that, as the dimensionless electron-phonon coupling constant $\lambda$ increases past an $O(1)$ value, the system tends to become unstable, either towards the formation of localized bipolarons or towards a charge density wave state. At the same time (and relatedly), the phonon frequency is renormalized downward as $\lambda$ increases, suppressing $T_c$.

Here, we construct an explicit $d-$dimensional model where these strong-coupling instabilities are avoided, and $T_c$ increases without bound upon increasing $\lambda$. The model includes $N$ electronic bands interacting with $N^2$ phonon modes. 
The model is solvable asymptotically in the large-$N$ limit; then, the famous Allen-Dynes result~\cite{Dynes1975} $T_c \propto \omega_0 \sqrt{\lambda}$ is valid for large $\lambda$, 

so long as $\lambda \ll N$,
and hence $T_c/\omega_0$ is unbounded as $N\to \infty$.
 Note that at the heuristic level, it is difficult to identify physical circumstances where more than a few phonons are comparably strongly coupled to the relevant electrons. Nevertheless, our analysis suggests that generically, the larger the number of phonon modes coupled to the electrons, the larger the $\lambda$ at which the suppression of $T_c$ onsets.
\end{enumerate}

\section{Flat band superconductivity:\\ bound on $T_c/T_F$?}
\label{sec:flat}

In most conventional superconductors, $T_c$ is determined by the energy scale associated with electron pairing. On the other hand, across numerous unconventional superconductors, $T_c$ is more strongly sensitive to the ``phase ordering scale"~\cite{emery1995importance}. In this context, an important recent advance is the result by Hazra, Verma, and Randeria (HVR)~\cite{Hazra18} of a rigorous upper bound on $\rho_s(T)$, the temperature-dependent superfluid phase stiffness, in terms of the integral of the optical conductivity over frequency (the optical sum rule). However, since this integral includes all the bands, this upper bound is often of the order of several electron-volts in electronic systems of interest.

At the heuristic level, this bound has been interpreted~\cite{park2021tunable} as implying a bound on $T_c/T_F$, where $T_F=E_F/k_B$ is the Fermi energy in units of temperature.\footnote{For a Galilean invariant system with a parabolic band, HVR express their bound in terms of the Fermi energy.} 

At the outset, it is important to define sharply what we mean by $E_F$. A particular protocol that is often adopted in experiments to estimate $E_F$ is to use an effective mass, $m^*$, obtained from quantum oscillations along with an estimate of the Fermi momentum ($k_F$) from a measurement of the the carrier density, and then defining $E_F = k_F^2/(2m^*)$. 
This procedure is only possible when there is a nearby Fermi liquid-like state that displays quantum oscillations. 

Below, we propose two different definitions of $E_F$, that we can use in settings that do not rely on any underlying assumptions (e.g., that there is a nearby Fermi liquid) and are also amenable to an experimental interpretation. We consider the case in which we add a given density, $n$, of electrons to an insulating reference state. We can define $E_F$ as follows:  
\begin{itemize}
\item Starting from our reference state, we set the temperature to zero and consider the change in the chemical potential, $\mu(n,T=0)$, as we fill in $n$ electrons,
\begin{equation}
E_F^{(1)} \equiv \mu(n,T=0) - \mu_0,
\label{eq:EF1}
\end{equation}
as an effective definition of $E_F$. Here $\mu_0 = \lim_{n\to 0} \mu(n,T=0)$. Note that the above definition of $E_F$ includes all many-body corrections, which can furthermore be dependent on the density itself, and does not make any reference to any non-interacting limit. 

\item Alternatively, we can define the Fermi energy through  the ``non-interacting" density of states, $\rho(\epsilon)$, for adding a single electron to the insulating system. In this case, the Fermi energy ($E_{F}^{(2)}$) is defined implicitly from the expression
\begin{equation}
n= \int_{\epsilon_{\mathrm{min}}}^{\epsilon_{\mathrm{min}}+E_{F}^{(2)}} d\epsilon\ \rho(\epsilon),
\label{eq:EF2}
\end{equation}
where $\epsilon_{\mathrm{min}}$ is the energy of the ground state with one electron added on top to the insulating reference state. $E_F^{(2)}$ is accessible directly in e.g. STM measurements \cite{Xie2019}.
\end{itemize}

\begin{figure}
	\begin{center}
          \includegraphics[width=0.99\columnwidth]{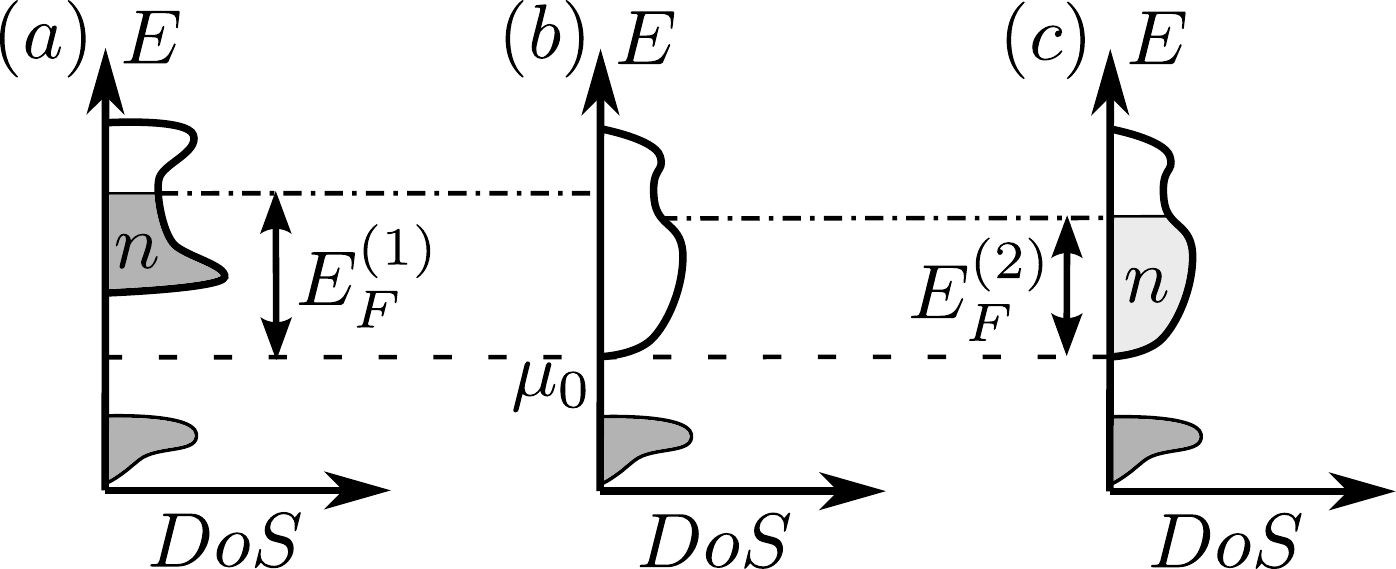} 
	\end{center}
	\caption{(a) 
	Illustration of the two operational definitions of $E_F$. (a,b) $E_F^{(1)}$ is defined in Eq. \eqref{eq:EF1} as the difference between the chemical potential at 
	density $n$ (panel a) and the limiting value of the chemical potential $\mu_0$ 
	at which $n$ approaches the value, $n_0$, it takes
	in a reference insulating state (panel b).
(c) $E^{(2)}_F$ is defined by Eq. \eqref{eq:EF2} in terms of the density of states $\rho(\varepsilon)$ of a single electron excitation 
added to the reference insulating state.}
	\label{fig:Efs}
\end{figure}

Below, we provide model Hamiltonians of interacting electrons in flat bands where the superconducting $T_c$ exceeds the Fermi energy by one or both of the above definitions. 
Thus, these models exemplify flat band superconductivity, where $T_c$ is determined entirely by the interaction scale~\cite{Shaginyan1990,Volovik2011,Volovik2013}. Analogous phenomena may also occur in semi-metals \cite{Kopnin2008,Kopnin2010}.

\subsection{Flat band superconductivity induced by a nearby dispersive band}
\label{pair}

We consider a model consisting of a nearly-flat band and a dispersive band.{\footnote{A closely related model \cite{LL21} has recently been studied in the context of superconductivity in twisted bilayer graphene. }}
The single-particle part of the Hamiltonian is given by
\begin{equation}
H_0 = \sum_{\k,\l,\sigma} \epsilon_\l(\k) c_{\k,\ell,\sigma}^\dagger c_{\k,\ell,\sigma}^{\vphantom{dagger}},
\label{eq:H0}
\end{equation}
where $c_{\k,\l,\sigma}^\dagger$ creates an electron with quasi-momentum $\k$ in band $\l=1,2$ and spin polarization $\sigma$. We consider the lower band ($\l=1$) to be a flat band with bandwidth, $W_1$, that we will ultimately take to be parametrically small (i.e. $W_1\rightarrow0$). The upper band ($\l=2$) is separated from the flat band by an energy gap, $\D_\tn{gap}$, and has a large bandwidth, $W_2\gg W_1$. The bands are topologically trivial and the Wannier functions are tightly localized on the lattice sites.

We now introduce an on-site interaction which scatters a pair of electrons between the flat band and the dispersive band:
\begin{subequations}
\beq
H_{\tn{int}} &=& V \sum_{\bm{R}}\left[ \phi_1^\dagger (\bm{R})\phi_2(\bm{R}) + \tn{h.c.}\right], \label{pairint} \\
\phi_\l(\r) &=& c_{\bm{R},\ell,\downarrow} c_{\bm{R},\ell,\uparrow},
\label{eq:Hint}
\eeq
\end{subequations}
where $\bm{R}$ labels a lattice site. 
Let us focus on the case where the flat band is half-filled, such that the number of particles per unit cell is $n=1$. 

We consider the case where $V\ll \Delta_{\rm{gap}}\ll W_2$. Within mean-field theory, the superconducting transition temperature is given by (see Appendix \ref{app:mft}),
\be
T_{\tn{MF}} \approx  \left[\frac {\nu_2 V^2}{8}\right]\ \ln \left[\frac {W_2}{\Delta_{\rm{gap}}}\right],
\label{Tmf}
\ee
where we have assumed a constant density of states per unit cell, $\nu_2 = 1/W_2$, in the dispersive band. The zero-temperature phase stiffness is given by (Appendix~\ref{app:mft})
\be
\rho_{s}(T=0) \approx \frac{V^2}{8\pi \Delta_{\rm{gap}}}.
\ee
Hence $\rho_s(0)\gg T_{\rm{MF}}$, which implies that phase fluctuations are unimportant in determining $T_c$~\cite{emery1995importance},  i.e. $T_c\approx T_{\rm{MF}}$.

We now examine the Fermi energies $E^{(1,2)}_F$ defined in Eqs.~(\ref{eq:EF1}) and (\ref{eq:EF2}) and compare them to $T_c$. Adding a single particle to the empty system, we find that $E_F^{(2)}\sim W_1 \ll T_c$, and hence $T_c/E_F^{(2)} $ can be made arbitrarily large.  $E_F^{(1)}$ is computed in Appendix~\ref{app:mft} by calculating the chemical potentials at $n=1$ and $n\rightarrow 0$. The result is $E_F^{(1)} = 2 T_{\rm{MF}} \sim T_c$. Hence, in this model, $T_c/E^{(1)}_F = O(1)$. An example of a different model where $T_c/E^{(1)}_F$ is unbounded is presented in the next section.

\subsection{Flat band superconductivity induced by spatial extent of Wannier functions}
\label{chiral}

We now introduce a different model for superconductivity in a narrow band. 
The model is defined on a two-dimensional square lattice with two electronic orbitals per unit cell. The Hamiltonian is given by
\begin{subequations}
\label{eq:chiral}
\begin{eqnarray}
    H &=& H_0 + H_U,\label{eq:modHam}\\
    H_0 & = & t \sum_{\substack{\k\\\gamma=\{\l,\sigma\}}} c^{\dagger}_{\k} \left(\tau_x\sin{\alpha_{\k}}  + \sigma_z \tau_y\cos{\alpha_{\k}}  -\mu  \right) c^{\phantom{\dagger}}_{\k}, \nn \\ 
    H_U &=& - \frac{U}{2} \sum_{\r,\l} \delta n^2_{\r,\l} + V\sum_{\langle \r,\r' \rangle,\l}  \delta n_{\r,\l} \delta n_{\r',\l}. 
\end{eqnarray}
\end{subequations}
Here, $c^\dagger_{\bm{k}} = (c^\dagger_{\bm{k},1,\uparrow},c^\dagger_{\bm{k},1,\downarrow},c^\dagger_{\bm{k},2,\uparrow},c^\dagger_{\bm{k},2,\downarrow})$, where the operator $c^{\dagger}_{\k,\ell,\sigma}$ creates an electron with momentum $\k$ in orbital $\l=1,2$ with spin $\sigma=\uparrow,\downarrow$. The Pauli matrices $\tau_{
x,y,z}$ and $\sigma_{
x,y,z}$ act on the orbital and spin degrees of freedom, respectively, and $\delta n_{\r,\l} \equiv \sum_{\sigma} c^{\dagger}_{\r,\l,\sigma} c^{\vphantom{\dagger}}_{\r,\l,\sigma} - 1$ is the number of particles at site $\r$ and orbital $\l$, relative to quarter filling ($n=1$). $\langle \r,\r' \rangle$ denotes nearest-neighbor sites. The single-particle Hamiltonian $H_0$ leads to perfectly flat bands with energies $\ve=\pm t$. The function $\alpha_{\k} \equiv \zeta (\cos{k_x}+\cos{k_y})$ controls the spatial extent of the Wannier functions in each band,  tuned by the dimensionless parameter $\zeta$.\footnote{More specifically, the Wannier functions decay exponentially over a distance proportional to $\zeta$.} Note that there is no obstruction towards constructing exponentially localized Wannier functions for the above model since the bands are topologically trivial.\footnote{This can be seen from the fact that the Berry curvature of the bands is identically zero, since $H_0$ contains only $\tau_{x,y}$ but not $\tau_z$.} The strength of on-site attractive Hubbard interaction is denoted $U>0$, whereas $V>0$ is a nearest-neighbor repulsion. 

We are interested in the limit where $T \ll U,\,V\ll \Delta_{\tn{gap}}~(=2t)$ and $\zeta\ll 1$.\footnote{An extensive numerical study of the model~\eqref{eq:chiral} beyond this parameter regime will appear in an upcoming publication.} In this limit, we project $H_{U}$ to the lower eigenband. The projected Hamiltonian is expanded in powers of $\zeta$. The average density is set to $n=1$ particles per unit cell.

For $\zeta=0$, the problem effectively reduces to a set of decoupled sites with a strong attractive interaction; the resulting ground state manifold is highly degenerate with local ``Cooper pairs'' but a vanishing phase stiffness. 
The linear corrections in $\zeta$ vanish due to a chiral symmetry and the orbital $l$ independent interaction strength $U$.
At second order in $\zeta$, the projected interaction, $\widetilde{H_U}$, contains nearest-neighbor pair-hopping and density-density interactions: 
\be
\widetilde{H_U} = - \sum_{\la\r,\r'\ra} \left[J_\perp(\hat{\eta}^x_{\bm{r}} \hat{\eta}^x_{\bm{r}'} + \hat{\eta}^y_{\r'} \hat{\eta}^y_{\r}) + J_z \hat\eta^z_{\r} \hat\eta^z_{\r'}\right].
\label{xxz}
\ee
Here, we have also introduced the pseudospin operators $\hat\eta^{a}_{\bm{r}}=(\Psi^\dagger_{\r} \eta^{a} \Psi^{\vphantom{\dagger}}_{\r})/2$, where $\Psi^\dagger_{\r}=(\tilde{c}^{\dagger}_{\bm{r},\uparrow}, \tilde{c}^{\phantom{\dagger}}_{\bm{r},\downarrow})$, 
$\tilde{c}^\dagger_{\r,\sigma}$ creates an electron with spin $\sigma$ in a Wannier-orbital localized around site $\r$ in the lower band of $H_0$ (with $\tau_y=-\sigma$ for $\zeta=0$), 
and $\eta^a$ are Pauli matrices. The $J_\perp$ and $J_z$ terms correspond to hopping and a nearest-neighbor interaction of the Cooper pairs, respectively, and their strengths are 
$J_\perp = \zeta^2(2U+V)/8$ and $J_z = \zeta^2 U/4 - V(2+9 \zeta^2/8)$.

For $V=0$, the system has an emergent $SU(2)$ symmetry that relates the pairing and charge order parameters~\cite{Tovmasyan2018}. This symmetry is weakly broken by terms of order $(U/t)^2$, not included in Eq.~\eqref{xxz}. For $0<V\ll U$, the problem is equivalent to a two-dimensional pseudospin ferromagnet with a weak easy-plane anisotropy. Parameterizing the anisotropy by $\Delta J = J_\perp - J_z$, we can estimate the critical temperature of the BKT transition as~\cite{Nelson1977}
\be
T_c \sim \pi J_\perp/\log(\pi J_\perp/\Delta J).
\label{eq:xxz_Tc}
\ee
Note that in the limit $\Delta J\rightarrow 0$ we get $T_c\rightarrow 0$, as required by the Mermin-Wagner-Hohenberg theorem. 

We now turn to estimating $E_F^{(1,2)}$. Due to the particle-hole symmetry of the effective Hamiltonian in Eq.~\eqref{xxz}, the chemical potential at $n=1$ (i.e., $\langle \hat{\eta}_{\r}^z\rangle = 0$) is $\mu(n=1)=0$. In the limit $n\rightarrow 0$, the system consists of dilute Cooper pairs. In this limit, the interactions between the Cooper pairs can be neglected, and the chemical potential is equal to half the energy per Cooper pair: $\mu(n\rightarrow 0) = -(J_\perp - J_z) = -\Delta J$.\footnote{Importantly, for $J_\perp>J_z$, the system does not phase separate at any density.} Therefore, $E_F^{(1)} = \mu(n=1) - \mu(n\rightarrow 0) = \Delta J$. Comparing this to Eq.~\eqref{eq:xxz_Tc}, we find that for $\Delta J \ll J_\perp$, $T_c \gg E_F^{(1)}$. Clearly, since the lower band is completely dispersionless, $E_F^{(2)}=0$. We conclude that $T_c$ can be made arbitrarily larger than the Fermi energy by either of the two definitions of Eqs.~(\ref{eq:EF1}) and (\ref{eq:EF2}). 

It is worth noting that, in the parameter regime we are considering, $\rho_s(0)\sim J_\perp \sim \zeta^2 U$. Hence, the delocalization of the Cooper pairs is entirely due to the interactions and the spatial overlap between the Wannier function of the active band, as in  Refs.~\cite{Torma15,Bernevig19,Rossi19,Torma19,JHEBDC,MR21}.{\footnote{The finite value of $\rho_s(0)$ and the associated lower bound as derived in Refs.~\cite{Torma15,Bernevig19,Rossi19,Torma19} is based on the application of BCS mean-field theory.}} Here, however, for $U\gg V$, we get $\Delta J\ll J_\perp$ and hence $\rho_s(0)\gg T_c$ [see Eq.~\eqref{eq:xxz_Tc}].

\begin{figure}
	\begin{center}
          \includegraphics[width=0.99\columnwidth]{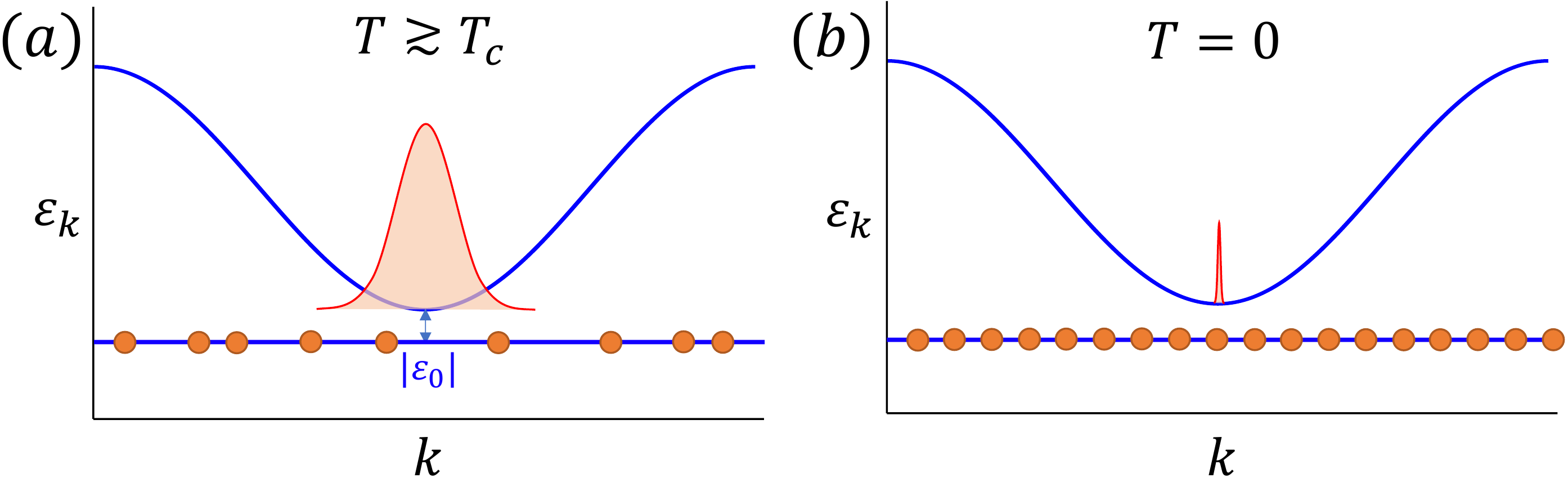} 
	\end{center}
	\caption{Schematic illustration of the dispersion and momentum distribution function of the bosons in the model with unbounded $T_c/\rho_s(0)$ [Eq.~\eqref{eq:Hb}]. (a) At $T\gtrsim T_c$, the flat band is approximately half filled, and the remaining $n_b-1/2$ bosonic particles per site occupy the dispersive band. (b) At $T=0$, the flat band is completely filled with one boson per site, and hence the superfluid density is proportional to $n_b-1$.}
	\label{fig:Tc_rhos}
\end{figure}

\section{Bound on $T_c/\rho_s(0)$?}
\label{sec:rhos}

In this section, we turn to the question of whether the zero temperature phase-stiffness, $\rho_s(0)$, sets an upper bound on $T_c$ in two dimensions. $\rho_s(0)$ can be extracted directly from a measurement of the London penetration depth ($\lambda_L^2(0)\propto 1/\rho_s(0)$), or from the imaginary part of the low-frequency optical conductivity.  

As is well known, in two spatial dimensions, the phase stiffness right below $T_c$ is related to $T_c$
by the inequality\footnote{At a continuous BKT transition, $\rho_s(T\rightarrow T^{-}_c)=2T_c/\pi$. However, if the transition is first order, $\rho_s$ right below $T_c$ can be larger than the universal BKT value. See, e.g., Ref. \cite{Domany1984}.} $\rho_s(T\rightarrow T^{-}_c)\geq 2T_c/\pi$.
However, $T_c$ is not directly related to $\rho_s(0)$. 
On physical grounds, it often makes sense to identify $\rho_s(0)$ with a ``phase ordering scale'' that sets an upper limit on $T_c$~\cite{emery1995importance}. This is justified by the fact that $\rho_s(T)$ is usually a monotonically decreasing function of temperature, i.e. $\rho_s(0)\geq \rho_s(T_c)$, and therefore $T_c$ can be bounded from above by $\rho_s(0)$. In conventional superconductors, $\rho_s(0)\gg T_c$, and $T_c$ is almost entirely determined by the pairing scale. In contrast, in underdoped cuprates, $\rho_s(0)$ is close to $T_c$, as illustrated by the famous Uemura plot~\cite{Uemura1989}. This suggests that in these systems, phase fluctuations play an important role in limiting $T_c$~\cite{emery1995importance}.

While this line of reasoning is likely correct in most circumstances, we will show here that  --- as a matter of principle ---  there
  is {\it no} bound on $T_c/\rho_s(0)$. We outline a concrete model where $\rho_s(0)$ can be made arbitrarily smaller than $\rho_s(T_c)$. 

Let us begin with a two-dimensional lattice model of two species of (complex) bosons, $b_1,~b_2$,
\begin{subequations}
\beq
\label{eq:Hb}
H^b &=& H^b_0 + H^b_{\tn{int}},\\
H^b_0 &=& \sum_{\alpha,\k} \ve_\alpha(\k)~ b_{\alpha,\k}^\dagger b^{\nd}_{\alpha,\k},
\eeq
\end{subequations}
where $\ve_2(\k)$ is assumed to have a large bandwidth, $W_2$, and $\ve_1 = \ve_2(0)-\ve_0$ forms a completely flat band at an energy $\ve_0$ below the bottom of the $\ve_2$ band, i.e. the $b_1$ bosons are completely localized on individual sites. The dispersion of the two species of bosons is illustrated in Fig.~\ref{fig:Tc_rhos}. For the purpose of our discussion here, we can approximate $\ve_2(\k)\approx k^2/2m_b$ near the bottom of the broad band. $H^b_{\tn{int}}$ includes an on-site (Hubbard) interaction of strength $U_{1,2}$ for the $b_{1,2}$ bosons. We take $U_2 \ll W_2$ whereas $U_1\rightarrow \infty$, implying that the number of $b_1$ bosons on each localized site can only be $0$ or $1$. The total average number of bosons per unit cell is chosen to be $n_b>1$. 

 At temperatures near $T_c$, the chemical potential is slightly above the bottom of the broad band. Then, assuming that $\ve_0 \ll T$, the average occupation of the localized sites is close to $1/2$ (since the $b_1$ bosons are essentially hard-core bosons at effectively `infinite' temperature), so there are approximately $n_b-1/2$ bosons per unit cell left to occupy the broad band. The critical temperature is $T_c \sim \frac{n_b-1/2}{2 m_b}$, up to logarithmic corrections in $\frac{n_b-1/2}{2 m_b}/U_2$ \cite{popov1972theory,kagan1987influence,Fisher1988}.\footnote{Since we are in two spatial dimensions, in the absence of interactions, $T_c=0$.} The momentum distribution of particles at $T\gtrsim T_c$ is shown schematically in Fig.~\ref{fig:Tc_rhos}a.

On the other hand, at $T=0$, all the localized sites are filled with one boson. The density of bosons in the broad band is thus $n_b-1$, and the superfluid stiffness is $\rho_s\approx \frac{n_b-1}{2 m_b}$. The boson distribution function is illustrated in Fig.~\ref{fig:Tc_rhos}b. Clearly, the ratio $T_c/\rho_s(0)$ can be made arbitrary large by letting $n_b\rightarrow 1^+$. If $1/2<n_b<1$, the ground state is not a superfluid, and there is a reentrant transition into a superconducting state with increasing $T$.

Note that 
in our simple model (Eq.~\eqref{eq:Hb}), the numbers of the two boson species are separately conserved. However, we do not expect 
the key results to be changed qualitatively 
by the addition of a weak hybridization between the two species, that breaks this separate conservation of the two boson numbers. 
In particular, a small hybridization 
generically produces a perturbative correction 
to $T_c$ and $\rho_s(0)$.

Indeed, a mild version of this sort of breakdown of the heuristic bound on $T_c/\rho_s(0)$ has been documented experimentally in Zn-doped cuprates \cite{uchida}. Here, the pristine material comes close to saturating the heuristic bound;  light Zn doping suppresses $T_c$ but apparently suppresses $\rho_s(0)$ more rapidly, leading to a ratio that slightly exceeds the value proposed in Ref. \cite{emery1995importance}. This was explained --- likely correctly --- by the authors of Ref. \cite{uchida} as being due to Zn-induced inhomogeneity of the superfluid response. This explanation is spiritually close to the  model discussed above: Each Zn impurity destroys the superconductor in its vicinity, possibly due to 
pinning of local spin-density-wave order \cite{hirota}. In some sense, this can be thought of as a state with localized d-wave pairs near the impurities. This effect depletes the condensate at low temperature, causing a decrease in the superfluid density. However, near $T_c$, this effect weakens, as the spin-density wave order partially melts.
From this perspective, it would be interesting to explore whether this violation can be made parametrically large with increasing Zn concentration - approaching the point at which $T_c\to 0$.

\section{Electron-phonon superconductivity: \\bound on $T_c/\omega_0$?}
\label{sec:el_ph}

Recently, it has been proposed that $T_c$ in an electron-phonon superconductor can never exceed a certain fraction of the characteristic phonon frequency, $\omega_0$~\cite{Esterlis2018,esterlis2018bound}. This putative bound implies that Migdal-Eliashberg (ME) theory~\cite{migdal1958interaction,eliashberg1960interactions,marsiglio2008electron} must break down when the dimensionless electron-phonon coupling $\lambda$ is of order unity~\cite{Esterlis2018,chubukov2020eliashberg}, since according to ME theory, $T_c$ grows without limit with increasing $\lambda$~\cite{Dynes1975}. 
In general, the failure of ME theory at $\lambda = O(1)$ is a result of strong-coupling effects: (i) The lattice tends to become unstable for large $\lambda$, resulting in a charge density wave (CDW) transition; (ii) electrons become tightly bound into bipolarons, whose kinetic energy is strongly quenched in the strong coupling limit; and (iii) as $\lambda$ increases, the phonon frequency renormalizes downward by an appreciable amount, $\Delta\omega$, suppressing $T_c$~\cite{Cohen06,note_softening}.

These strong-coupling effects certainly play an important role in limiting $T_c$ in real systems, where it is typically found never to exceed about $0.1\,\omega_0$ across numerous conventional superconductors~\cite{esterlis2018bound}. Determinant Monte Carlo simulations of the paradigmatic Holstein model reveal that ME theory indeed fails for $\lambda=O(1)$, and the maximal $T_c$ is significantly below $\omega_0$~\cite{Esterlis2018}. As we shall now show, however, this is not a rigorous bound on $T_c$.  

To demonstrate this, we consider a particular large$-N$ variant of the electron-phonon problem~\cite{Werman2016,werman2017non} with $N-$component electrons and $N\times N-$component (`matrix') phonons, defined on a $d-$dimensional hypercubic lattice. The Hamiltonian is given by
\begin{subequations}
\beq
H &=& H_e + H_{\tn{ph}} + H_{\tn{e-ph}}, \label{elph}\\
H_e &=& \sum_{a=1}^{N}\sum_{\r,\r',\sigma}(-t_{\r\r'}-\mu\delta_{\r\r'})c_{\r,\sigma,a}^{\dagger}c^{\vphantom{\dagger}}_{\r',\sigma,a},\\
H_{\tn{ph}} &=& \sum_{a,b=1}^{N}\sum_{\r}\left(\frac{P_{\r,ab}^{2}}{2M}+\frac{K}{2}X_{\r,ab}^{2}\right),\\
H_{\tn{e-ph}} &=& \frac{\alpha}{\sqrt{N}}\sum_{a,b=1}^{N}\sum_{\r,\sigma}X^{\vphantom{\dagger}}_{\r,ab}c_{\r,\sigma,a}^{\dagger}c^{\vphantom{\dagger}}_{\r,\sigma,b}.
\label{eq:alpha}
\eeq
\end{subequations}
Here, $c^\dagger_{\r,\sigma,a}$ creates an electron at position $\r$ with spin $\sigma$ 
in ``orbital'' $a$. The hopping parameters $t_{\r\r'}$ and chemical potential $\mu$ are assumed to be identical for all orbitals. We have introduced a real, symmetric matrix of phonon displacements, $\hat{X}_{\r}$ and their canonically conjugate momenta, $\hat{P}_{\r}$, with frequency $\omega_{0}=\sqrt{K/M}$, assumed to be much smaller than the Fermi energy. The phonons are taken to be dispersionless for simplicity. The purely on-site electron-phonon coupling constant is denoted $\alpha$, with a $N-$dependent normalization factor that ensures that the model has a finite energy density in the $N\rightarrow\infty$ limit. The dimensionless electron-phonon coupling constant is defined as $\lambda = \alpha^2 \nu(0)/K$, where $\nu(0)$ is the electronic density of states at the Fermi level per orbital.  

\begin{figure}
	\begin{center}
          \includegraphics[width=0.99\columnwidth]{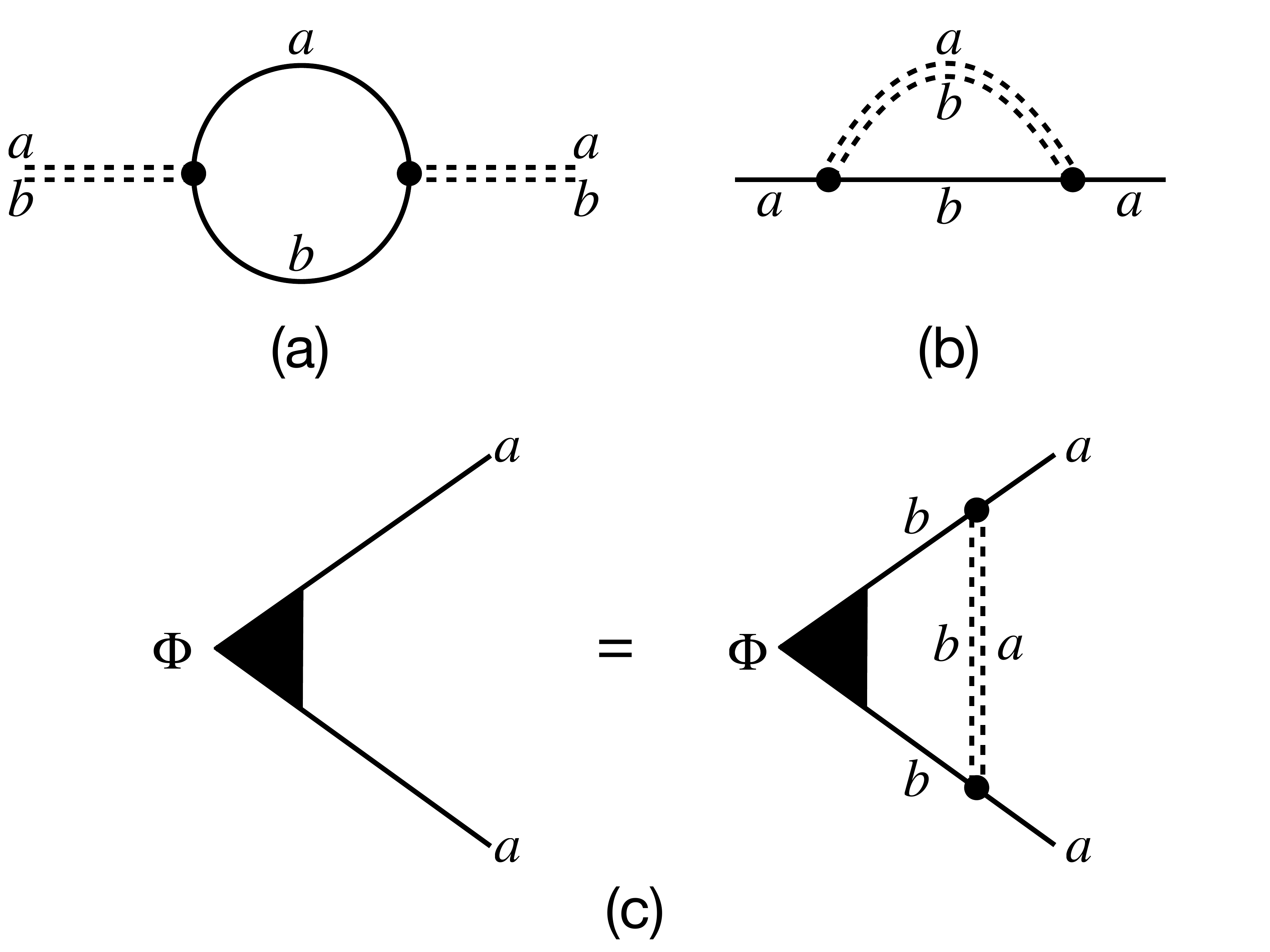} 
	\end{center}
	\caption{Self-energy for the (a) phonon ($X_{ab}$), and (b) electron ($c_a$) fields, respectively. (c) Bethe-Salpeter equation for the pairing vertex, $\Phi$. Double dashed and solid lines represent the phonon and electron fields. Solid circles (triangles) denote the electron-phonon (pairing) vertex.}
	\label{fig:diags}
\end{figure}

We are interested in the large$-N$ limit of the model defined in Eq.~\eqref{elph}. 
Since the number of phonon degrees of freedom is much larger than the number of electron orbitals, the phonon dynamics are essentially unaffected by the coupling to the electrons, even when the electrons are strongly perturbed.  
This implies that the strong-coupling effects mentioned above are suppressed, even for 
$\lambda \gg 1$.  In particular, as we show in Appendix~\ref{app:stable},
there is no lattice instability or polaron formation as long as $\lambda \ll N$, and the softening of the phonon frequency is only of the order of $\Delta\omega\sim \lambda \omega_0/N$. 

To zeroth order in $1/N$, the equations for the electron self-energy and the pairing vertex are exactly those given by Eliashberg theory, whereas the phonon self-energy is of order $1/N$ (see Fig.~\ref{fig:diags}). Thus, the self-consistent equations for the pairing vertex are identical to those of ME theory neglecting the renormalization of the phonons, and hence the result is the same. In particular, for $1 \ll \lambda \ll N$, $T_c \approx 0.1827 \,\omega_0 \sqrt{\lambda}$~\cite{Dynes1975}. 
Implicit in the fact that Migdal-Eliashberg theory is exact at $N\to \infty$ is an assurance that there is no suppression of $T_c$ by phase fluctuations.\footnote{In the $d=2$ version of our model, the BKT temperature differs from the mean-field transition temperature only by a $1/N$ correction.}   More specifically, this follows from the observation that the superfluid stiffness is ${\cal O}(N)$.
Thus, $T_c$ is unbounded.  

The key ingredient in our model that allows us to take $\lambda>1$ without suffering from lattice instabilities is that the different phonon modes couple to electron bilinears that do not commute with each other [see Eq.~\eqref{eq:alpha}]. This limits the energy gain from distorting a given set of phonon modes when forming a CDW or a polaron bound state, since the resulting perturbations to the electronic Hamiltonian cannot be diagonalized simultaneously. In contrast, the contributions of the individual phonon modes add algebraically in the total dimensionless coupling $\lambda$ that enters the equation for the pairing vertex (the same dimensionless coupling also determines the resistivity in the normal state of this model~\cite{Werman2016}). 

It is worth noting that while these considerations may be of some  
use in searching for systems with ever higher $T_c$, as a practical matter it may be difficult to significantly violate the proposed heuristic bound.  To achieve  $T_c \approx \omega_0$ requires 
 the extremely large value of $\lambda\approx 25$.  
 At the same time, to avoid polaron formation requires $N \gg \lambda$, which means the number of distinctly coupled phonon modes would have to be $N^2 \gg \lambda^2 \sim 625$!

\section{Outlook}
\label{outlook}

The notion of a fundamental upper bound on $T_c$ for models of interacting electrons is an attractive concept. In this paper, we have demonstrated that while there are numerous physical settings where such bounds can be formulated at a heuristic level, there exists no fundamental, universal upper bound on $T_c$ in terms of the characteristic energy scales of interest to us, which include an appropriately defined $T_F$, $\rho_s(0)$ and $\omega_0$. We have constructed explicit counter-examples which violate these heuristic bounds by an arbitrary amount. 

On the experimental front, it would be  fruitful to look for candidate materials where the heuristic bounds are violated by a large amount. 
The fact that these bounds are usually satisfied is to be expected, since although the bounds are not rigorous, the physical reasoning behind them is quite robust. 
As our theoretical discussion illustrates, whenever such bounds are violated, there is an interesting underlying physical reason behind the violation; moreover, the mechanisms behind the violation of the heuristic bounds may suggest ways to optimize $T_c$. Our work provides two such examples. Flat band systems with a large spatial extent of the Wannier functions are a promising platform for increasing $T_c$. In electron-phonon systems, the instabilities that limit $T_c$ at large electron-phonon interaction strength can be partially mitigated if the coupling is shared between several active phonon modes that couple to non-commuting electronic operators.

\acknowledgements
We thank Pablo Jarillo-Herrero, Mohit Randeria, and J.M. Tranquada for stimulating discussions. SAK was supported, in part, by the National Science Foundation (NSF) under Grant No. DMR2000987. DC was supported by the faculty startup grants at Cornell University. EB and JH were supported by the European Research Council (ERC) under grant HQMAT (Grant Agreement No. 817799), the Israel-US Binational Science Foundation (BSF), and by a Research grant from Irving and Cherna Moskowitz.

\appendix

\section{Additional details for flat band superconductivity induced by nearby dispersive band}
\subsection{Mean-field gap equations}
\label{app:mft}
Here, we derive the expressions for $T_{{\rm MF}}$, $\rho_{s}(0)$,
and $E^{(1)}_{F}$ for the model in Eqs. (3,4) of the main text. 

We replace the Hamiltonian of Eqs. (3,4) by the mean-field Hamiltonian
\begin{equation}
    H_{\rm{MF}} = H_0 + \sum_{\bm{R},\ell} \Delta_\ell^{\vphantom{\dagger}} c^\dagger_{\bm{R},\ell\uparrow}c^\dagger_{\bm{R},\ell\downarrow} + \rm{h.c.}
\end{equation}
The mean-field equations for the pairing potentials $\Delta_{\ell=1,2}$ are given by
\begin{equation}
\Delta_{1}=V\Delta_{2}\int\frac{d^{2}k}{\left(2\pi\right)^{2}}\frac{\tanh\left(\beta\sqrt{\xi_{2}^{2}(\bm{\bm{k}})+|\Delta_{2}|^{2}}/2\right)}{2\sqrt{\xi_{2}^{2}(\bm{\bm{k}})+|\Delta_{2}|^{2}}},\label{eq:MF1}
\end{equation}
\begin{equation}
\Delta_{2}=V\Delta_{1}\frac{\tanh\left(\beta\sqrt{\xi_{1}^{2}+|\Delta_{1}|^{2}}/2\right)}{2\sqrt{\xi_{1}^{2}+|\Delta_{1}|^{2}}}.\label{eq:MF2}
\end{equation}
Here, we have set the width of the first band to zero, such that
\begin{equation}
\xi_{1}=-\mu,
\end{equation}
whereas, for simplicity, we have taken the dispersion of the second
band to be parabolic with an effective mass $m$: 
\begin{equation}
\xi_{2}(\bm{\bm{k}})=\Delta_{\mathrm{gap}}+\frac{k^{2}}{2m}-\mu.
\end{equation}
The total width of this band is $W_{2}$.

The density of particles per unit cell is 
\begin{equation}
n=\sum_{\ell=1,2}\int\frac{d^{2}k}{\left(2\pi\right)^{2}}\left(1-\frac{\xi_{\ell}(\bm{\bm{k}})}{E_{\ell}(\bm{\bm{k}})}\tanh\left[\frac{\beta E_{\ell}(\bm{\bm{k}})}{2}\right]\right),\label{eq:n}
\end{equation}
where $E_{\ell}(\bm{k})=\sqrt{\xi_{\ell}^{2}(\bm{\bm{k}})+\Delta_{\ell}^{2}}$. We set the density to $n=1$. 

We first solve for the critical temperature $T_{{\rm MF}}$. We assume
that $\mu\approx0$, to be checked self-consistently. We then linearize
Eqs. (\ref{eq:MF1},\ref{eq:MF2}) in $\Delta_{1,2}$ at $\beta=1/T_{{\rm MF}}$.
Eq. (\ref{eq:MF2}) gives $\Delta_{2}=V\Delta_{1}/(4T_{{\rm MF}})$;
inserting this in Eq. (\ref{eq:MF1}) gives $T_{{\rm MF}}=V^{2}\log\left(W_{2}/\Delta_{\mathrm{gap}}\right)/(8W_{2})$.
Under our assumptions, we get that $T_{{\rm MF}}\ll\Delta_{{\rm gap}}$, and therefore
the number of particles in the second band is exponentially small
(setting $\Delta_{2}=0$ at $T=T_{{\rm MF}}$). Hence, Eq.~(\ref{eq:n})
indeed gives that $\mu\approx0$ corresponds to a filling of $n=1$,
consistently with our starting assumption. 

Next, we compute $E_{F}^{(1)}$. The chemical potential at $n\rightarrow0$
can be obtained by solving the linearized gap equations at $T=0$:
\begin{align}
\Delta_{1} & =\frac{V\Delta_{2}}{2}\int\frac{d^{2}k}{\left(2\pi\right)^{2}}\frac{1}{\left|\xi_{2}(\bm{\bm{k}})\right|},\\
\Delta_{2} & =V\Delta_{1}\frac{1}{2|\mu|},
\end{align}
which give the chemical potential $\mu$ at which superconductivity
onsets. This gives $\mu(n\rightarrow0)=-2T_{{\rm MF}}$. To find the
chemical potential at $n=1$ and $T=0$, we solve the gap equations
at $T=0$. Assuming that $\mu(n=1)\approx0$, we obtain $\Delta_{2}=V/2$,
$\Delta_{1}=V^{2}\log\left(W_{2}/\Delta_{{\rm gap}}\right)/\left(4W_{2}\right).$
Inserting this back into Eq. (\ref{eq:n}) gives $\mu(n=1)\propto|\Delta_{1}||\Delta_{2}|/\sqrt{W_{2}\Delta_{\rm gap}}\sim V^{3}/(W_{2}^{3/2}\Delta_{\rm gap}^{1/2})\ll\mu(n\rightarrow0)$,
justifying our assumption that $\mu$ is small. Thus we find that
$E_{F}^{(1)}=\mu(n=1)-\mu(n\rightarrow0)=2 T_{{\rm MF}}$, the result
quoted in the text. 

\subsection{Phase stiffness}
\label{app:stiffness}

Let us compute the phase stiffness at $T=0$ within mean-field theory. We couple the system to an external gauge field, and differentiate the ground state energy twice with respect to vector potential. At $T=0$, the entire contribution to the stiffness arises from the diamagnetic term. Using $\Delta_2= V/2 \ll\Delta_{\rm{gap}}$, we can further expand the result to leading order in $|\Delta_2|$ such that,
\beq
\rho_s(0) &=& \frac{V^2}{m} \int_\omega \int_{\k} \frac{-1}{(i\omega-\xi_2(\k))^2 (-i\omega - \xi_2(\k))}\nn\\
 &=& \frac{V^2}{8\pi \Delta_{\tn{gap}}}.
\eeq

Note that, within mean-field theory, we can write the average particle number in the $\l=2$ band as
\beq
\langle n_{\l=2}\rangle = \int \frac{d^2 k}{(2\pi)^2} \bigg(1 - \frac{\xi_2(\k)}{\sqrt{\xi_2^2(\k) + |\Delta_2|^2}} \bigg),
\eeq
which can be similarly expanded to leading order in a small $|\Delta_2|$ as,
\beq
\langle n_{\l=2}\rangle \approx \int \frac{d^2 k}{(2\pi)^2}  \frac{|\Delta_2|^2}{2\xi_2^2(\k)} = 2m\rho_s(0).
\eeq
Thus, in the particular flat-band limit for $\l=1$, we find that the phase stiffness is determined by the usual Galilean invariant result, $\rho_s(0) = n_s/m$, where $n_s$ now corresponds to the average particle density in the dispersive $\l=2$ band whose mass is $m$.

\section{Additional details for flat band superconductivity induced by spatial extent of Wannier functions}
\label{app:chiral}


First, we diagonalize the single-particle Hamiltonian of Eq.~\eqref{eq:chiral} and find the flat eigenbands at energy $\epsilon_\pm=\pm t$ and their wave functions $\phi_\pm(l,\bk,\sigma)=\frac{1}{\sqrt{2}} e^{i\sigma\alpha_\bk\tau_z/2} (1,\pm\sigma i)^T$. It is convenient to introduce $d^{\dagger}_{\bk,\sigma}$, the creation operator for the target band $\epsilon_-$ with momentum $\bk$ and spin $\sigma$, as well as its Fourier-transformed real-space operator $d^{\dagger}_{\br_i,\sigma}=\sum_{l,\br} \psi_{\br_i,\sigma}(l,\br) \, c^{\dagger}_{\br,l,\sigma}$.
The wave function $\psi_{\br_i,\sigma}(l,\br_j)=\frac{1}{A} \sum_{\bk} e^{-i(\br_i-\br_j)\bk} \phi_-(l,\bk,\sigma)$ is tightly localized around $\br_i$. In fact, $\psi_{\br_i,\sigma}(l,\br_j)\sim (i\zeta)^{|\delta_x|+|\delta_y|} + ...$ decays exponentially with distance ${\bf \delta} = \br_j-\br_i$ and for $\zeta=0$,  $\psi_{\br_i,\sigma}(l,\br_j)=\frac{\delta_{ij}}{\sqrt{2}} (1,\,\sigma i)^T$.

This allows to decompose the microscopic operators in the band basis, $c^{\dagger}_{l,\br_i,\sigma} = \sum_{\br} \tilde{\psi}_{l,\br_i,\sigma}(\br) \, d^{\dagger}_{\br,\sigma} +\dots$, where we have omitted contributions from the remote bands, as well as the interaction term,
\begin{widetext}
\begin{eqnarray}
    c^{\dagger}_{l,\br_i,\sigma}c^{\phantom{\dagger}}_{l,\br_i,\sigma} c^{\dagger}_{l,\br_j,\sigma'}c^{\phantom{\dagger}}_{l,\br_j,\sigma'} &=& \sum_{\br_1\cdots \br_4} \tilde{\psi}_{l,\br_i,\sigma}(\br_1) \tilde{\psi}^*_{l,\br_i,\sigma}(\br_2)
    \tilde{\psi}_{l,\br_j,\sigma'}(\br_3) \tilde{\psi}^*_{l,\br_j,\sigma'}(\br_4) \, 
    d^{\dagger}_{\br_1,\sigma} d^{\phantom{\dagger}}_{\br_2,\sigma} 
    d^{\dagger}_{\br_3,\sigma'} d^{\phantom{\dagger}}_{\br_4,\sigma'} + \dots 
    \, ,\label{eq:proj_int}
\end{eqnarray}
\end{widetext}
with $\tilde{\psi}_{l,\br_i,\sigma}(\br_j)=\frac{1}{A} \sum_{\bk} e^{-i(\br_i-\br_j)\bk} \phi_-^{*}(l,\bk,\sigma)$. Note that $\alpha_{\bk}=\alpha_{-\bk}$ such that $\tilde{\psi}_{l,\br_i,\sigma}(\br_j)=\psi_{\br_i,\sigma}(l,\br_j)^*$.

As mentioned in the main text, we are interested in the limit where $T \ll U,\,V\ll \Delta_{\tn{gap}}~(=2t)$ and $\zeta\ll 1$. For $V=0$ and $\zeta=0$, we recover the Hubbard interaction with a renormalized coupling strength $U \sum_l|\psi_{\br_i}(l,\br_i)|^4 = U/2$. Hence, each site $\br_i$ is either empty or double occupied at low temperatures, $T\ll U$. 

For finite $\zeta$, one might expect correlated hopping terms at linear order. However, those terms vanish identically. Due to the absence of $\tau_z$ in Eq.~\eqref{eq:chiral} the system has a chiral symmetry, which guaranties that the wavefunctions of the low-energy band have equal weights on both orbitals. This, combined with the orbital-independent coupling strength $U$, implies that all terms linear in $\zeta$ originating from orbital $l=1$ are canceled by $l=2$ contributions. Note that such a hopping term, e.g., $d^{\dagger}_{\br,\sigma} d^{\phantom{\dagger}}_{\br,\sigma}  d^{\dagger}_{\br,\sigma'} d^{\phantom{\dagger}}_{\br+\hat{x},\sigma'}$, necessarily generates two single-occupied sites and hence creates an excitation with an energy of order $U$. Accordingly, terms linear in $\zeta$ cannot show up in the projected Hubbard interaction as leading order perturbation even if the chiral symmetry is weakly broken or if the coupling strength $U$ was weakly orbital dependent.

Hence, the leading order corrections are quadratic in $\zeta$. Nearest neighbor density interaction and pair hopping terms preserve the local parity and thus are relevant perturbations within the $\zeta=0$ ground-state sector. Their effective coupling strength are given by $ \sum_l|\psi_{\br_i}(l,\br_i)|^2|\psi_{\br_i}(l,\br_j)|^2$ and $ \sum_l \psi^*_{\br_i}(l,\br_i)^2 \psi_{\br_i}(l,\br_j)^2$, respectively, up to combinatorial factors. 

\section{Additional details of the electron-phonon model}
\label{app:elph}

\subsection{Stability in the large-$N$ limit}
\label{app:stable}
We study the stability of the system towards CDW order and polaron formation at $T=0$ in
the limit $M\rightarrow\infty$ keeping $K$ fixed, i.e., $\omega_{0}=\sqrt{K/M}\rightarrow0$.
We assume that setting $M$ finite or $T>0$ will only make the system more stable, as these promote quantum and thermal fluctuations of the phonons, respectively. 

To second order in the phonon displacements, the ground state energy of the system
can be computed by perturbation theory. The result is
\begin{equation}
E=\sum_{\bm{q},a,b}K\left(\frac{1}{2}-\frac{\lambda}{N}\frac{\chi_{\bm{q}}}{\chi_{\bm{0}}}\right)\left|X_{\bm{q},ab}\right|^{2},\label{eq:E}
\end{equation}
where $\chi_{\bm{q}}$ is the static charge susceptibility: 
\begin{equation}
\chi_{\bm{q}}=\frac{2}{L^{d}}\sum_{\bm{k}}\frac{\Theta(\varepsilon_{\bm{k}+\bm{q}})-\Theta(\varepsilon_{\bm{k}})}{\varepsilon_{\bm{k}+\bm{q}}-\varepsilon_{\bm{k}}},
\end{equation}
where $L$ is the linear dimension of the system. The condition for stability with respect to CDW formation is that the term in parenthesis in Eq. (\ref{eq:E}) be non-negative for all $\bm{q}$. Note the $1/N$ prefactor of the second term in Eqn.~ \ref{eq:E}. Hence,
in the large-$N$ limit, the system does not possess a CDW instability. By the same token, polaron formation is also suppressed, since a local deformation of the phonons accompanied by an electronic bound state is never favored. These instabilities only appear when $\lambda$ becomes of the order of $N$.

\subsection{Eliashberg equation}
\label{app:eliashberg}

The Eliashberg equations for electron-phonon superconductivity and
their solution at large $\lambda$ are quite standard~\cite{Dynes1975,marsiglio2008electron,chubukov2020eliashberg}. For completeness,
we will nevertheless outline the derivation below.

In the large-$N$ limit, the electron self-energy at $T>T_{c}$ obeys the following self-consistent equation, depicted in Fig.~\ref{fig:diags}b:
\begin{align}
\Sigma(i\omega_{n}) & =\alpha^{2}T\sum_{\bm{k}',\omega_{n}'}\frac{1}{i\omega'_{n}-\varepsilon_{\bm{k}'}-\Sigma(i\omega'_{n})}D_0(i\omega_n-i\omega_n')\nonumber \\
 & =-i\pi\lambda T\sum_{\omega_{n}'}\frac{\omega_{0}^{2}\,{\rm sgn}(\omega'_{n})}{\omega_{0}^{2}+\left(\omega_{n}-\omega'_{n}\right)^{2}},
 \label{eq:Sigma}
\end{align}
where $D_0(i\Omega_n)$ is the {\it bare} phonon Green's function
\be
D_0(i\Omega_n) = \frac{1}{K+M\Omega_n^{2}}.
\ee
The phonon self-energy, shown in Fig.~\ref{fig:diags}a, is of order $1/N$ in our model, and is thus neglected.
In the second line of Eq.~\eqref{eq:Sigma}, we have performed the $\bm{k}'$ summation, assuming
that at the frequencies of interest, $W\gg\left|\omega'_{n}-\Sigma(i\omega'_{n})\right|$,
where $W$ is the electronic bandwidth. At temperatures large compared
to $\omega_{0}$, the largest contribution to the self-energy comes
from $\omega_{n}=\omega'_{n}$, giving $\Sigma(i\omega_{n})\approx-i\pi\lambda T{\rm sgn}(\omega_{n})$.

The linearized Eliashberg equation for the pairing vertex $\Phi(i\omega_{n})$, represented diagrammatically in Fig.~\ref{fig:diags}c,
is given by
\begin{align}
\Phi(i\omega_{n}) & =\alpha^{2}T\sum_{\bm{k}',\omega_{n}'}\frac{\Phi(i\omega'_{n})}{\left|i\omega'_{n}-\varepsilon_{\bm{k}'}-\Sigma(i\omega'_{n})\right|{}^{2}}D_0(i\omega_n-i\omega_n')\nonumber \\
 & =\pi\lambda T\sum_{\omega_{n}'}\frac{\Phi(i\omega'_{n})}{\left|i\omega'_{n}-\Sigma(i\omega'_{n})\right|}\frac{\omega_{0}^{2}}{\omega_{0}^{2}+\left(\omega_{n}-\omega'_{n}\right)^{2}}.\label{eq:Phi}
\end{align}
It is convenient to define the Eliashberg $Z$ factor: 
\begin{equation}
Z(i\omega_{m})=\frac{i\omega_{n}-\Sigma(i\omega_{m})}{i\omega_{n}},
\end{equation}
and the gap function $\Delta(i\omega_{n})=\Phi(i\omega_{n})/Z(i\omega_{n})$.
Using Eq.~\eqref{eq:Sigma}, one can rewrite Eq.~(\ref{eq:Phi}) as
\begin{align}
\Delta(i\omega_{n}) & =\pi\lambda T\sum_{\omega_{n}'}\frac{\Delta(i\omega'_{n})-\Delta(i\omega{}_{n})\frac{\omega'_{n}}{\omega_{n}}}{\left|\omega'_{n}\right|}\frac{\omega_{0}^{2}}{\omega_{0}^{2}+\left(\omega_{n}-\omega'_{n}\right)^{2}}.
\end{align}
In the limit of large $\lambda$, $T_{c}$ is much larger than $\omega_{0}$.
Importantly, the $\omega_{n}=\omega'_{n}$ term in the sum above contribution
vanishes, which allows us to take the limit $\omega_{0}/T\rightarrow0$
in the deminator of the right-hand side. This results in the following
equation:
\begin{align}
\Delta(n) & =\frac{\lambda\omega_{0}^{2}}{\left(2\pi T\right)^{2}}\sum_{n'\ne n}\frac{\Delta(n')-\Delta(n)\frac{2n'+1}{2n+1}}{\left|2n'+1\right|}\frac{1}{\left(n-n'\right)^{2}}.\label{eq:eliashberg_strong}
\end{align}
This equation has a solution when
\begin{equation}
\frac{\lambda\omega_{0}^{2}}{T_{c}^{2}}=c,
\end{equation}
where $c=O(1)$. 

A numerical solution of Eq.~(\ref{eq:eliashberg_strong}) gives~\cite{Dynes1975}
\begin{equation}
T_{c}=0.1827\,\omega_{0}\sqrt{\lambda}.
\end{equation}

\bibliographystyle{apsrev4-1_custom}
\bibliography{references}
\end{document}